\documentclass[a4paper]{jpconf}
\usepackage{graphicx}
\pdfoutput=1
\usepackage{pstricks}
\usepackage{color}
\usepackage{amssymb,amsmath,bbm}
\usepackage{epsf}
\usepackage{epsfig}
\usepackage{afterpage}
\usepackage{longtable}
\usepackage{latexsym,mathrsfs,dsfont}
\usepackage{graphics}
\usepackage{url}
\usepackage{paralist}
\usepackage{bbold}

\newcommand{\tev}{\, {\rm TeV}}
\newcommand{\gev}{\, {\rm GeV}}

\newcommand{\vcb}{|V_{cb}|}

\newcommand{\vub}{|V_{ub}|}

\newcommand{\ord}{\mathcal{O}}
\newcommand{\bsi}{B_6^{(1/2)}}
\newcommand{\bei}{B_8^{(3/2)}}

\def\epe{\varepsilon'/\varepsilon}
\newcommand{\be}{\begin{equation}}
\newcommand{\ee}{\end{equation}}

\def\kpn{K^+\rightarrow\pi^+\nu\bar\nu}

\def\klpn{K_{L}\rightarrow\pi^0\nu\bar\nu}

\newcommand{\kepe}{\kappa_{\varepsilon^\prime}}
\newcommand{\keps}{\kappa_{\varepsilon}}
\def\eps{\varepsilon}
\newcommand{\UonePr}{{\mathrm{U(1)_{L_\mu-L_\tau}}}}

\begin{document}
\title{ Kaon Flavour Physics Strikes Back}

\author{ Andrzej~J.~Buras}

\address{TUM Institute for Advanced Study, Lichtenbergstr. 2a, D-85748 Garching, Germany}

\ead{aburas@ph.tum.de}

\begin{abstract}
 In this short presentation I emphasize the increased importance of kaon flavour physics in the search for new physics (NP) that we should witness in the rest of this decade and in the next decade. The main actors will be 
the branching ratios for the rare decays $\kpn$ and $\klpn$, to be measured by NA62 and KOTO, and their correlations with  the ratio $\epe$ on which recently 
progress by lattice QCD and large $N$ dual QCD approach has been made implying 
a new flavour anomaly. Further correlations of $\kpn$, $\klpn$ and $\epe$ with 
$\varepsilon_K$, $\Delta M_K$, $K_L\to\mu^+\mu^-$ and $K_L\to\pi^0\ell^+\ell^-$ will help us to identify indirectly possible NP at short distance scales.
This talk summarizes the present highlights of this facinating field including 
some results from concrete NP scenarios. To be published online by the Institute of Physics Proceedings.
\end{abstract}

\section{Introduction}
In three recent reports \cite{Buras:2015hna,Buras:2016qia,Buras:2016bch} I have stressed the increased importance of kaon flavour physics in the search for new physics (NP) which we should witness in the 
near future. Indeed after years of silence I expect that kaon flavour physics 
will strike back providing new insights in the dynamics at very short distance 
scales. The following pages can be considered as an express review of this 
fascinating field. Further details and in particular numerous references 
can be found in \cite{Buras:2015hna,Buras:2016qia,Buras:2016bch,Buras:2015nta}.

\section{Important Messages}\label{sec:2}
\subsection{$\epe$}\label{sub:2.1}
Presently in kaon flavour physics the most exciting appears to be the 
anomaly in $\epe$. The present status of $\epe$ in the SM  can be summarized as follows. The
RBC-UKQCD lattice collaboration calculating hadronic matrix elements of 
all contributing operators but not including isospin breaking effects finds 
\cite{Blum:2015ywa, Bai:2015nea}
\begin{align}
  \label{eq:epe:LATTICE}
  (\epe)_\text{SM} & = (1.38 \pm 6.90) \times 10^{-4},\qquad {\rm (RBC-UKQCD)}.
\end{align}
Using the hadronic matrix elements of QCD-penguin ($Q_6$) and EW-penguin ($Q_8$)
$(V-A)\otimes (V+A)$ operators from RBC-UKQCD lattice collaboration 
but extracting the matrix elements
of penguin $(V-A)\otimes (V-A)$ operators from the data on CP-conserving $K\to\pi\pi$ 
amplitudes and including isospin breaking effects one finds \cite{Buras:2015yba}
\begin{align}
  \label{eq:epe:LBGJJ}
  (\epe)_\text{SM} & = (1.9 \pm 4.5) \times 10^{-4},\qquad {\rm (BGJJ)}\,
\end{align}
that is confirmed within the errors by the recent analysis  in \cite{Kitahara:2016nld}
\begin{align}
  \label{KNT}
  (\epe)_\text{SM} & = (1.1 \pm 5.1) \times 10^{-4},\qquad {\rm (KNT)}\,.
\end{align}

All these results are  significantly below the experimental world
average from NA48 \cite{Batley:2002gn} and KTeV \cite{AlaviHarati:2002ye, 
Abouzaid:2010ny} collaborations, 
\begin{align}
  \label{eq:epe:EXP}
  (\epe)_\text{exp} & = (16.6 \pm 2.3) \times 10^{-4} ,
\end{align}
suggesting that models providing enhancement of $\epe$ are favoured. 

These results are based on NLO calculations of the Wilson coefficients 
of the relevant operators \cite{Buras:1991jm,Buras:1992tc,Buras:1992zv,Ciuchini:1992tj,Buras:1993dy,Ciuchini:1993vr}.
Partial NNLO calculations have been performed in \cite{Buras:1999st,Gorbahn:2004my,Brod:2010mj}. Complete 
NNLO result from Maria Cerda-Sevilla, Martin Gorbahn, Sebastian J{\"a}ger and Ahmet Kokulu should be available soon.

While these results, based on the hadronic matrix elements from RBC-UKQCD 
lattice collaboration, suggest some evidence for NP in $\epe$,
the large uncertainties in the hadronic matrix elements in question do not yet
preclude that eventually the SM will agree with data. Therefore the 
upper bounds on the relevant hadronic matrix elements of $Q_6$ and $Q_8$
from large $N$ dual QCD approach \cite{Buras:2015xba} are important as they give presently the strongest support to the anomaly 
in question, certainly stronger than present lattice results. 

In the strict large $N$ limit \cite{Buras:1985yx,Bardeen:1986vp,Buras:1987wc} the parameters $\bsi$ and $\bei$ 
that represent the relevant hadronic matrix elements of the QCD penguin operator $Q_6$ 
and the electroweak penguin operator $Q_8$, respectively, are simply given by
\be\label{LN}
\bsi=\bei=1, \qquad {\rm (large~N~Limit)}\,.
\ee
But RBC-UKQCD results \cite{Bai:2015nea,Blum:2015ywa} imply
\cite{Buras:2015yba,Buras:2015qea}
\be\label{Lbsi}
\bsi=0.57\pm 0.19\,, \qquad \bei= 0.76\pm 0.05\,, \qquad (\mbox{RBC-UKQCD}),
\ee
and the  suppression of $\bsi$ below unity 
is the main origin of the strong suppression of $\epe$ below the data within the SM. Yet in view of the large error in $\bsi$ one could 
be sceptical about any claims that there is 
NP in $\epe$. Future lattice results could in principle raise $\bsi$ towards its large $N$ value and above $\bei$ bringing the SM result for $\epe$ close
to its experimental value. 

However, the analyses of $\bsi$ and $\bei$ within the dual QCD approach in
\cite{Buras:2015xba,Buras:2016fys} show that such a situation is rather 
unlikely. Indeed, in this approach going beyond the strict large $N$ limit  
one can understand the suppression of $\bsi$ and $\bei$ below the unity 
 as the effect of the meson 
evolution from scales  $\mu=\ord(m_\pi,m_K)$  at which (\ref{LN}) is valid to 
 $\mu=\ord(1\gev)$ at which Wilson coefficients of $Q_6$ and $Q_8$ are 
evaluated \cite{Buras:2015xba}. This evolution has to be matched to the usual perturbative quark evolution for scales higher than $1\gev$ and in fact the supressions in question
and the property that $\bsi$ is more strongly suppressed than $\bei$ are 
consistent with the perturbative evolution of these parameters above  
$\mu=\ord(1\gev)$. Thus we are rather confident that \cite{Buras:2015xba}
\be\label{NBOUND}
\bsi< \bei < 1 \, \qquad ({\rm dual~QCD}).
\ee
 For further details, see \cite{Buras:2015xba,Buras:2016qia}.

Additional support for the small value of $\epe$ in the SM comes from the recent reconsideration of the role of final state interactions (FSI) in $\epe$ 
\cite{Buras:2016fys}. Already long time ago the chiral perturbation theory practitioners put forward the idea
that both the amplitude ${\rm Re}A_0$, governed by the current-current operator  $Q_2-Q_1$ and the $Q_6$ contribution to the ratio $\epe$ could be 
enhanced significantly through FSI in a correlated manner (see e.g. \cite{Pallante:2001he} and other reference in \cite{Buras:2016qia}).
However, as shown recently in \cite{Buras:2016fys}
 FSI are likely to be important for the $\Delta I=1/2$  rule, in agreement with
these papers but much less relevant  for $\epe$.

It should finally be noted that 
even without  lattice results, varying all input parameters, the bound 
in (\ref{NBOUND}) implies the upper bound on $\epe$ in the SM
 \be\label{BoundBGJJ}
(\epe)_\text{SM}<(8.6\pm 3.2) \times 10^{-4} \,,\qquad {(\rm BG)}\,.
\ee
On the other hand employing the lattice value for $\bei$ in (\ref{Lbsi}) and
$\bsi=\bei=0.76$, one obtains  $(6.0\pm 2.4)\times 10^{-4}$  instead of (\ref{BoundBGJJ}), well below the data. 

All these findings give strong motivation for searching for NP which could enhance $\epe$ above its SM value. We will summarize the present efforts in this 
direction below.
\subsection{Tensions between $\varepsilon_K$ and $\Delta M_{s,d}$ in the SM and CMFV Models}\label{sub:2.2}
In \cite{Blanke:2016bhf} we have pointed out a significant tension between $\varepsilon_K$ and $\Delta M_{s,d}$ within the SM
 and models with constrained MFV (CMFV) implied by 
 new lattice QCD results from Fermilab Lattice and MILC Collaborations \cite{Bazavov:2016nty}   on $B^0_{s,d}-\bar B^0_{s,d}$ hadronic matrix elements. 
Even if this tension is certainly not as large as is the case of 
the $\epe$ anomaly 
the  plots in \cite{Blanke:2016bhf}, in particular in Fig.~5
of that paper, show that there is a clear tension between $\varepsilon_K$ and $\Delta M_{s,d}$ in the SM and CMFV models. Moreover this tension persists independently of the values of CKM parameters. For smaller (exclusive) values of $\vcb$ one finds $\Delta M_{s,d}$  to agree well with the data, while  $\varepsilon_K$ is roughly $25\%$ below its experimental value. For  $\vcb$ in the 
ballpark of inclusive determinations one finds $\varepsilon_K$ to agree with
the data, while  $\Delta M_{s,d}$ are then typically by $15\%$ larger than their experimental values.  These numbers are for the SM, in all other CMFV models the situation
gets worse.  

The improved $\Delta B=2$ hadronic matrix 
elements from other lattice collaborations and  improved 
values of  $\vcb$ and $\vub$  will tell us one day whether this tension 
persists and if this will turn out to be the case, whether there is a
$\varepsilon_K$  anomaly and/or a $\Delta M_{s,d}$ anomaly.
\subsection{$\kpn$ and $\klpn$ in the SM}
These two rare decays allow to test the short distance scales far beyond the 
reach of the LHC. Even scales of $\ord(100)\tev$ can be probed in this manner \cite{Buras:2014zga}.
The present status of $\kpn$ and $\klpn$ within the SM  has been 
presented in \cite{Buras:2015qea} with the result 
\begin{align}\label{PREDA}
    \mathcal{B}(\kpn) &= \left(8.4 \pm 1.0\right) \times 10^{-11}, \\
    \mathcal{B}(\klpn) &= \left(3.4\pm 0.6\right) \times 10^{-11}.    
\end{align}
But the most 
important outcome of this paper are 
 parametric expressions for the branching ratios of these two 
decays in terms of the CKM input and the correlations between   $\kpn$ and 
$B_{s}\to\mu^+\mu^-$  and  between $\kpn$ and $\varepsilon_K$ in the SM. These 
formulae should allow to  monitor the numerical values for these branching ratios within the SM when the CKM input improves. Interesting correlations 
between $\kpn$ and $\klpn$ and various observables are found in simplified models with flavour violating couplings of the SM $Z$ and of a heavy $Z^\prime$ \cite{Buras:2015yca}.

\subsection{Strategy for $\epe$ and Lessons}\label{sec:strategy}
In 
order to investigate the implications of $\epe$ anomaly on rare decays $\kpn$ and $\klpn$ in a systematic fashion a strategy has been proposed in \cite{Buras:2015jaq}. While $\epe$ plays the dominant role in this strategy it was useful 
to assume that there is also a modest $\varepsilon_K$ anomaly.

Then $\epe$ and 
$\varepsilon_K$ in the presence of NP contributions are given by
\be\label{GENERAL}
\frac{\varepsilon'}{\varepsilon}=\left(\frac{\varepsilon'}{\varepsilon}\right)^{\rm SM}+\left(\frac{\varepsilon'}{\varepsilon}\right)^{\rm NP}\,, \qquad \varepsilon_K\equiv 
e^{i\varphi_\eps}\, 
\left[\varepsilon_K^{\rm SM}+\varepsilon^{\rm NP}_K\right] \,
\ee
with NP contributions parametrized as follows:
\be\label{deltaeps}
\left(\frac{\varepsilon'}{\varepsilon}\right)^{\rm NP}= \kepe\cdot 10^{-3}, \qquad   0.5\le \kepe \le 1.5, \qquad
\varepsilon_K^{\rm NP}= \keps\cdot 10^{-3},\qquad 0.1\le \keps \le 0.4 \,.
\ee
The ranges for $\kepe$ and $\keps$ indicate the required size of this contribution but can be kept as free parameters. They will be determined one day when the 
theory on $\epe$ and the CKM input improve.

In the simplest NP scenarios with tree-level $Z$ and $Z^\prime$ exchanges, 
the imaginary parts of flavour-violating $Z$ or $Z^\prime$ couplings to quarks 
are then determined as functions of $\kepe$. As $\varepsilon_K$ is governed by the product of imaginary and real parts of these couplings, invoking it  allows then to determine the corresponding real parts as functions of $\kepe$ and $\keps$.

 Having fixed the flavour violating couplings of $Z$ or $Z^\prime$ 
in this manner, one can express NP contributions to the branching ratios for $\kpn$, $\klpn$, $K_L\to\mu^+\mu^-$
and to $\Delta M_K$ in terms of $\kepe$ and $\keps$. 
Explicit formulae can be found in \cite{Buras:2015jaq}. In this manner one 
can directly study the impact of $\epe$ and $\varepsilon_K$ anomalies in $Z$ and $Z^\prime$ scenarios on these four observables.  

In \cite{Buras:2015jaq} numerous plots for the ratios
\be\label{Rnn+}
R^{\nu\bar\nu}_+\equiv\frac{\mathcal{B}(\kpn)}{\mathcal{B}(\kpn)_\text{SM}},\qquad
R^{\nu\bar\nu}_0\equiv \frac{\mathcal{B}(\klpn)}{\mathcal{B}(\klpn)_\text{SM}}\,
\ee
as functions of $\kepe$ and $\keps$ within the models with 
tree-level $Z$ and $Z^\prime$ exchanges have been presented. We will list the most 
important lessons from this study that depend on the
flavour violating couplings 
$\Delta^{sd}_{L,R}(Z)$ and $\Delta^{sd}_{L,R}(Z^\prime)$ \cite{Buras:2012jb}.
Moreover, we will use 
the abbreviations: $({\rm LHS\, \equiv\, left-handed~scenario})$ and $({\rm RHS\, \equiv\, right-handed~scenario})$
for NP scenarios in which only left-handed (LH) or right-handed (RH) flavour-violating couplings 
are present. The first six lesson deal with tree-level $Z$ exchanges, the remaining four with $Z^\prime$ tree-level exchanges.

{\bf Lesson 1:}
In the LHS, a given request for the enhancement of $\epe$ determines 
the coupling ${\rm Im} \Delta_{L}^{s d}(Z)$.

{\bf Lesson 2:}
In LHS an enhanced $\epe$ implies uniquely  {\it suppression} of
  $\mathcal{B}(\klpn)$. This property is known from NP scenarios in which 
 NP to $\klpn$ and $\epe$ enters dominantly through the modification of 
$Z$-penguins. 

{\bf Lesson 3:} 
The imposition of the $K_L\to\mu^+\mu^-$ constraint in LHS 
determines the range for
${\rm Re} \Delta_{L}^{s d}(Z)$ which with the already fixed ${\rm Im} \Delta_{L}^{s d}(Z)$ allows to calculate the shifts in $\varepsilon_K$ and $\Delta M_K$. 
They are very small. 

{\bf Lesson 4:}
With fixed ${\rm Im} \Delta_{L}^{s d}(Z)$ and the allowed range 
for ${\rm Re} \Delta_{L}^{s d}(Z)$, the range for  $\mathcal{B}(\kpn)$ 
can be obtained. Both an enhancement and a suppression of  $\mathcal{B}(\kpn)$ are possible. $\mathcal{B}(\kpn)$  can be enhanced by a factor of $2$ at most.

{\bf Lesson 5:}
Analogous pattern is found in RHS, although the numerics is different. See
  Fig.~1 in \cite{Buras:2015jaq}. In particular
the suppression of $\mathcal{B}(\klpn)$ for a given $\kepe$ is smaller.
Moreover,  an enhancement of $\mathcal{B}(\kpn)$  up to a factor of  $5.7$ 
is possible.

{\bf Lesson 6:}
In a general $Z$ scenario with LH and RH flavour-violating couplings the pattern of NP effects 
changes because 
LR operators dominate NP contributions to  $\varepsilon_K$ and $\Delta M_K$. 
One can then enhance
simultaneously $\epe$, $\varepsilon_K$, $\mathcal{B}(\kpn)$ and $\mathcal{B}(\klpn)$ which is not possible in LHS and RHS.  The correlations 
between $\epe$ and $\kpn$ and $\klpn$ depend sensitively on the ratio of 
real and imaginary parts of the flavour-violating couplings involved. 
Moreover large departures from 
SM predictions for $\kpn$ and $\klpn$ are possible and $\epe$ anomaly can be explained.

$Z^\prime$ models 
exhibit quite different pattern of NP effects in the $K$ meson system than the 
LH and RH $Z$ scenarios. In $Z$ scenarios only electroweak 
penguins (EWP) can contribute to $\epe$ in an important manner because of flavour dependent diagonal $Z$ coupling to quarks. But in $Z^\prime$ models the diagonal quark couplings can be flavour universal so that QCD penguin operators (QCDP) can dominate NP contributions to $\epe$. Interestingly,
the pattern of NP in rare $K$ decays depends on whether NP in $\epe$ is dominated by QCDP  or EWP operators.  

As demonstrated  in \cite{Buras:2015jaq} there
is a large hierarchy between real and imaginary parts of the  flavour 
violating couplings implied by $\epe$ anomaly in  QCDP and EWP scenarios.
In  the case of QCDP  imaginary parts dominate over the real ones, while
in the case of EWP this hierarchy is opposite  unless
the $\varepsilon_K$ anomaly is absent. Because of these different patterns there are  striking differences in the implications of the $\epe$ anomaly
for the correlation between $\kpn$ and $\klpn$ in these
two NP  scenarios if significant NP contributions to $\epe$ are required. 
The plots in \cite{Buras:2015jaq}  and in 
particular analytic derivations presented there illustrate these 
differences in a spectacular manner. The main lessons are as follows. 

{\bf Lesson 7:}
In the case of QCDP scenario the correlation between 
$\mathcal{B}(\klpn)$ and $\mathcal{B}(\kpn)$ takes place along the branch 
parallel to the Grossman-Nir bound  \cite{Grossman:1997sk}.

{\bf Lesson 8:}
In the EWP scenario the correlation   between 
$\mathcal{B}(\klpn)$ and $\mathcal{B}(\kpn)$  
is very different 
from the one of the QCDP case. NP effects in rare $K$ decays
 turn out to be modest in this case unless the diagonal quark couplings are $\ord(10^{-2})$ and then the requirement of 
shifting upwards $\epe$ implies large effects in $\kpn$ and $\klpn$ also in the EWP scenario. 

{\bf Lesson 9:}
For fixed values of the neutrino and  diagonal quark couplings in $\epe$ the 
predicted enhancements of $\mathcal{B}(\klpn)$ and $\mathcal{B}(\kpn)$ 
are much larger when NP in QCDP is required to remove the 
$\epe$ anomaly than it is the case of EWP. 

{\bf Lesson 10:}
 In QCDP scenario  $\Delta M_K$ is {\it suppressed} and this 
effect increases with increasing  $M_{Z^\prime}$ whereas in the EWP scenario 
 $\Delta M_K$ is {\it enhanced} and this effect decreases with increasing
 $M_{Z^\prime}$ as long as real couplings dominate.  Already on the basis of this property one could differentiate between 
these two scenarios when the SM prediction for $\Delta M_K$ improves.

\section{Results in specific NP models}\label{sec:NP}
\subsection{Preliminaries}
The  latest analyses of NP  contributions to $\epe$ in models with tree-level $Z$ and $Z^\prime$ exchanges like 331 models, Littlest Higgs model with T-parity  can be found in \cite{Blanke:2015wba,Buras:2015yca,Buras:2015kwd,Buras:2015jaq,Buras:2016dxz}.
The analyses in supersymmetric models can be found in \cite{Tanimoto:2016yfy,Kitahara:2016otd,Endo:2016aws}. In view of space limitations we will only briefly 
summarize the results in 331 models and models with vector-like quarks.

\subsection{331 Flavour News}\label{sec:331}
The 331 models are based on the gauge group $SU(3)_C\times SU(3)_L\times U(1)_X$ \cite{Singer:1980sw,Pisano:1991ee,Frampton:1992wt,Foot:1992rh,Montero:1992jk}.
In these models new contributions to $\epe$ and other flavour observables are  dominated by tree-level exchanges of a $Z^\prime$ with non-negligible contributions from tree-level $Z$ exchanges generated through the $Z-Z^\prime$ mixing. The size of these NP effects depends on $M_{Z^\prime}$,  on a parameter $\beta$, which distinguishes between various 331 
models, on fermion representations under the gauge group and a
parameter $\tan\bar\beta$ present in the $Z-Z^\prime$ mixing. Extensive recent
analyses in these models can be found in  \cite{Buras:2012dp,Buras:2013dea,Buras:2014yna,Buras:2015kwd,Buras:2016dxz}. References to earlier analyses of flavour physics in 331 models can be found there and in  \cite{Diaz:2004fs,CarcamoHernandez:2005ka}.

A detailed analysis of 331 models with different values of 
$\beta$, $\tan\bar\beta$ for two fermion representations $F_1$ and $F_2$, with 
the third SM quark generation belonging respectively to an antitriplet and a triplet under 
the $SU(3)_L$, has been presented in \cite{Buras:2014yna}: 24 models in total.
Requiring that these models perform
at least as well as the SM, as far as electroweak tests are concerned, seven models have been selected 
for a more detailed study of FCNC processes. 
Recent updated 
analyses of these seven models, that address the $\epe$ anomaly, have been presented in \cite{Buras:2015kwd,Buras:2016dxz} and we 
summarize the main results of these two papers. 
The main findings of \cite{Buras:2015kwd,Buras:2016dxz} for $M_{Z^\prime}=3\tev$ are as follows:
\begin{itemize}
\item
Among seven 331 models singled out through electroweak precision study only three  (M8, M9, M16) can provide significant shift of $\epe$ but not larger than $6\times 10^{-4}$, that is $\kepe\le 0.6$.
\item
The tensions between $\Delta M_{s,d}$  and $\varepsilon_K$ can be removed in these models.
\item
Two of them (M8 and M9) can simultaneously suppress $B_s\to\mu^+\mu^-$ and 
  bring the theory within $1\sigma$ range of
the combined result from CMS and LHCb. The most recent result from ATLAS \cite{Aaboud:2016ire}, while not accurate, appears to confirm this picture. On the other hand these models 
do not really help in the case of $B_d\to K^*\mu^+\mu^-$ anomalies \cite{ Altmannshofer:2014rta,Descotes-Genon:2015uva}.
\item
In M16 the situation is opposite. The rate for $B_s\to\mu^+\mu^-$ can be reduced for  $M_{Z^\prime}=3\tev$ by only a small amount  but 
 the anomaly  in $B_d\to K^*\mu^+\mu^-$ can be significantly reduced.
\item
For higher values of  $M_{Z^\prime}$ the effects in $B_s\to\mu^+\mu^-$ and $B_d\to K^*\mu^+\mu^-$ are small.
 NP effects in rare $K$ decays and $B\to K(K^*)\nu\bar\nu$ remain small in all 331 models even for  $M_{Z^\prime}$ of a few TeV. This could be challenged by NA62, KOTO and Belle II experiments in this decade.
\end{itemize}

All these results are valid for $\vub=0.0036$. For its inclusive value 
of $\vub=0.0042$, we find that for  $\vcb=0.040$ the maximal shifts in $\epe$ are increased to 
 $7.7 \times 10^{-4}$ and $8.8\times 10^{-4}$ for  $M_{Z^\prime}=3\tev$ and 
 $M_{Z^\prime}=10\tev$, respectively. Renormalization group effects are responsible for this enhancement of $\epe$ for increased  $M_{Z^\prime}$. A recent analysis in the MSSM in \cite{Kitahara:2016otd}
identifies this effect as well. But as explained in  \cite{Buras:2015kwd} 
 eventually for very high  $M_{Z^\prime}$, NP effects in $\epe$ will be suppressed.

Thus the main message from \cite{Buras:2015kwd,Buras:2016dxz} is that NP 
contributions in 331 
models can simultaneously solve $\Delta F=2$ tensions, enhance $\epe$ and 
suppress either the rate for $B_s\to\mu^+\mu^-$ or $C_9$ Wilson coefficient
without any significant NP effects on $\kpn$ and $\klpn$ and $b\to s\nu\bar\nu$
transitions. While sizable NP effects in $\Delta F=2$ observables and 
$\epe$  can persist for $M_{Z^\prime}$ outside the reach of the LHC, such 
effects in $B_s\to\mu^+\mu^-$ will only be detectable provided $Z^\prime$ 
will be discovered soon. 

\subsection{Models with vector-like quarks(VLQs)}\label{sec:VLQs}
A detailed analysis of flavour violation patterns in the $K$ and $B_{s,d}$
sectors in eleven models with VLQs has been presented in \cite{Bobeth:2016llm}. The simplest (five of them) are the ones in which the gauge group is the SM one and the only new particles are VLQs in a single complex representation under the 
SM gauge group. A general classification of such models and references to the rich literature can be found in \cite{Bobeth:2016llm,Ishiwata:2015cga}. In these models $\Delta F=1$ FCNCs are dominated by tree-level $Z$ exchanges, while $\Delta F=2$ transitions by box diagrams with VLQs and scalars provided $M_\text{VLQ}\ge 5\tev$. Otherwise tree-level $Z$ contributions cannot be neglected.

The  summary of patterns of flavour violation in these models 
can be found in  three DNA tables (Tables 5, 6, 10 in  \cite{Bobeth:2016llm}) and the numerical results in Tables 8 and 9 of that paper. Our extensive numerical analysis has shown that NP effects in several of these models can 
still be very large and that simultaneous consideration of several flavour observables should allow to distinguish between these models. In particular models with 
left-handed and right-handed flavour violating currents can be distinguished from each other in this manner. Here we list most important results of this paper.
\begin{itemize}
\item
All tensions between $\Delta M_{s,d}$ and $\varepsilon_K$ can be easily removed 
in these models because the usual CMFV correlations between  $\Delta M_{s,d}$ and $\varepsilon_K$ are not valid in them. The box diagrams with 
VLQs and Higgs scalar exchanges are dominantly responsible for it.
\item
Tree-level $Z$ contributions to $\epe$ can be large so that significant upward 
shift in $\epe$ can easily be obtained bringing the theory to agree with data.
\item
Simultaneously the branching ratio for $\kpn$ can be significantly enhanced over
its SM prediction, but only in models with flavour violating RH
  currents. In models with only LH currents $\kpn$ branching ratio can have at most its  SM value because of the $K_L\to\mu\bar\mu$ constraint. On the other hand the 
positive shift in $\epe$ implies uniquely suppression of the $\klpn$ branching ratio with the suppression being smaller in models with RH currents. The fact
that in models with RH currents $\kpn$ can be enhanced, while $\klpn$ suppressed is a clear signal of non-MFV sources at work. But also in models with pure LH currents 
the correlations between the branching ratios of these two decays differ from the MFV one.
\item
These  features distinguish VLQ-models  from 331 models,  discussed 
above, in which NP effects are dominated 
by $Z^\prime$ exchanges with the maximal shift in $\epe$ amounting to $0.8\times 10^{-3}$ and NP effects in rare $K$ decays being very 
small.
\item
Significant suppressions of the branching ratio for $B_s\to\mu^+\mu^-$, in particular in models with LH currents are possible.
While such effects are also possible in 331 models, they cannot be as large
as in VLQ models.
\item
On the other hand while 331 models can provide solutions to some LHCb anomalies,
this is not possible in  the VLQs models with SM gauge group and future confirmation of these anomalies
could turn out to be a problem for the latter models.
\end{itemize}

Having the latter possibility in mind we have considered also six  VLQ models with a heavy $Z^\prime$  related to $\UonePr$ symmetry and extended scalar sector. Some of such models,
considered already in \cite{Altmannshofer:2014cfa}, 
can explain LHCb anomalies     but NP effects in other observables are in my view less interesting 
than in models based on the SM gauge group. We refer to  \cite{Bobeth:2016llm}
for details.
Future experimental results on $\kpn$, $\klpn$, $B_s\to\mu^+\mu^-$ and 
LHCb anomalies and improved theoretical results on $\epe$ will tell us 
which of these VLQ models, if any, is selected by nature.

While the discovery of VLQs at the LHC would give a strong impetus to the 
models considered by us, non-observation of them at the LHC would not preclude 
their importance for flavour physics. In fact we have shown that large NP
effects in flavour observables can be present for  $M_{\rm VLQ} = 10$~TeV and 
in the flavour precision era one could even be sensitive to higher masses. 
 In this context we have pointed out that
the combination of $\Delta F=2$ and $\Delta F=1$ observables in a given
meson system allows to determine the masses of VLQs in a given representation
independently of the size of Yukawa couplings.

In summary the future of kaon flavour physics looks great and the coming years should be very exciting. 

\section*{Acknowledgements}
 It is 
a pleasure to thank Monika Blanke, Fulvia De Fazio and Jean-Marc G{\'e}rard for  very 
efficient studies discussed in this write-up and Christoph Bobeth, Alejandro Celis 
and Martin Jung for a very extensive analysis of VLQ models.
I would also like to thank the organizers of Kaon 2016 for inviting me 
to this very interesting conference  and for an impressive hospitality.
This research was done and financed in the context of the ERC Advanced Grant project ``FLAVOUR''(267104). It was also partially
supported by the DFG cluster of excellence ``Origin and Structure of the Universe''.

\section*{References}

\bibliographystyle{iopart-num}
\bibliography{allrefs}

\providecommand{\newblock}{}
\begin{thebibliography}{10}
\expandafter\ifx\csname url\endcsname\relax
  \def\url#1{{\tt #1}}\fi
\expandafter\ifx\csname urlprefix\endcsname\relax\def\urlprefix{URL }\fi
\providecommand{\eprint}[2][]{\url{#2}}

\bibitem{Buras:2015hna}
Buras A~J 2015 {\em PoS\/} {\bf EPS-HEP2015} 602 (\textit{Preprint}
  \eprint{1510.00128})

\bibitem{Buras:2016qia}
Buras A~J 2016 (\textit{Preprint} \eprint{1606.06735})

\bibitem{Buras:2016bch}
Buras A~J 2016 {\em {8th International Workshop on QCD - Theory and Experiment
  (QCD@Work 2016) Martina Franca, Italy, June 27-30, 2016}\/}
  (\textit{Preprint} \eprint{1609.05711})

\bibitem{Buras:2015nta}
Buras A~J 2015 {\em PoS\/} {\bf FWNP} 003 (\textit{Preprint}
  \eprint{1505.00618})

\bibitem{Blum:2015ywa}
Blum T {\em et~al.\/} 2015 {\em Phys.~Rev.\/} {\bf D91} 074502
  (\textit{Preprint} \eprint{1502.00263})

\bibitem{Bai:2015nea}
Bai Z {\em et~al.\/} (RBC, UKQCD) 2015 {\em Phys. Rev. Lett.\/} {\bf 115}
  212001 (\textit{Preprint} \eprint{1505.07863})

\bibitem{Buras:2015yba}
Buras A~J, Gorbahn M, J{\"a}ger S and Jamin M 2015 {\em JHEP\/} {\bf 11} 202
  (\textit{Preprint} \eprint{1507.06345})

\bibitem{Kitahara:2016nld}
Kitahara T, Nierste U and Tremper P 2016  (\textit{Preprint}
  \eprint{1607.06727})

\bibitem{Batley:2002gn}
Batley J {\em et~al.\/} (NA48) 2002 {\em Phys.~Lett.\/} {\bf B544} 97--112
  (\textit{Preprint} \eprint{hep-ex/0208009})

\bibitem{AlaviHarati:2002ye}
Alavi-Harati A {\em et~al.\/} (KTeV) 2003 {\em Phys.~Rev.\/} {\bf D67} 012005
  (\textit{Preprint} \eprint{hep-ex/0208007})

\bibitem{Abouzaid:2010ny}
Abouzaid E {\em et~al.\/} (KTeV) 2011 {\em Phys. Rev.\/} {\bf D83} 092001
  (\textit{Preprint} \eprint{1011.0127})

\bibitem{Buras:1991jm}
Buras A~J, Jamin M, Lautenbacher M~E and Weisz P~H 1992 {\em Nucl.~Phys.\/}
  {\bf B370} 69--104

\bibitem{Buras:1992tc}
Buras A~J, Jamin M, Lautenbacher M~E and Weisz P~H 1993 {\em Nucl.~Phys.\/}
  {\bf B400} 37--74 (\textit{Preprint} \eprint{hep-ph/9211304})

\bibitem{Buras:1992zv}
Buras A~J, Jamin M and Lautenbacher M~E 1993 {\em Nucl.~Phys.\/} {\bf B400}
  75--102 (\textit{Preprint} \eprint{hep-ph/9211321})

\bibitem{Ciuchini:1992tj}
Ciuchini M, Franco E, Martinelli G and Reina L 1993 {\em Phys. Lett.\/} {\bf
  B301} 263--271 (\textit{Preprint} \eprint{hep-ph/9212203})

\bibitem{Buras:1993dy}
Buras A~J, Jamin M and Lautenbacher M~E 1993 {\em Nucl.~Phys.\/} {\bf B408}
  209--285 (\textit{Preprint} \eprint{hep-ph/9303284})

\bibitem{Ciuchini:1993vr}
Ciuchini M, Franco E, Martinelli G and Reina L 1994 {\em Nucl.~Phys.\/} {\bf
  B415} 403--462 (\textit{Preprint} \eprint{hep-ph/9304257})

\bibitem{Buras:1999st}
Buras A~J, Gambino P and Haisch U~A 2000 {\em Nucl.~Phys.\/} {\bf B570}
  117--154 (\textit{Preprint} \eprint{hep-ph/9911250})

\bibitem{Gorbahn:2004my}
Gorbahn M and Haisch U 2005 {\em Nucl.~Phys.\/} {\bf B713} 291--332
  (\textit{Preprint} \eprint{hep-ph/0411071})

\bibitem{Brod:2010mj}
Brod J and Gorbahn M 2010 {\em Phys.~Rev.\/} {\bf D82} 094026
  (\textit{Preprint} \eprint{1007.0684})

\bibitem{Buras:2015xba}
Buras A~J and Gerard J~M 2015 {\em JHEP\/} {\bf 12} 008 (\textit{Preprint}
  \eprint{1507.06326})

\bibitem{Buras:1985yx}
Buras A~J and G\'erard J~M 1986 {\em Nucl.~Phys.\/} {\bf B264} 371

\bibitem{Bardeen:1986vp}
Bardeen W~A, Buras A~J and G\'erard J~M 1986 {\em Phys.~Lett.\/} {\bf B180} 133

\bibitem{Buras:1987wc}
Buras A~J and G\'erard J~M 1987 {\em Phys.~Lett.\/} {\bf B192} 156

\bibitem{Buras:2015qea}
Buras A~J, Buttazzo D, Girrbach-Noe J and Knegjens R 2015 {\em JHEP\/} {\bf 11}
  033 (\textit{Preprint} \eprint{1503.02693})

\bibitem{Buras:2016fys}
Buras A~J and Gerard J~M 2016  (\textit{Preprint} \eprint{1603.05686})

\bibitem{Pallante:2001he}
Pallante E, Pich A and Scimemi I 2001 {\em Nucl. Phys.\/} {\bf B617} 441--474
  (\textit{Preprint} \eprint{hep-ph/0105011})

\bibitem{Blanke:2016bhf}
Blanke M and Buras A~J 2016 {\em Eur. Phys. J.\/} {\bf C76} 197
  (\textit{Preprint} \eprint{1602.04020})

\bibitem{Bazavov:2016nty}
Bazavov A {\em et~al.\/} (Fermilab Lattice, MILC) 2016 {\em Phys. Rev.\/} {\bf
  D93} 113016 (\textit{Preprint} \eprint{1602.03560})

\bibitem{Buras:2014zga}
Buras A~J, Buttazzo D, Girrbach-Noe J and Knegjens R 2014 {\em JHEP\/} {\bf
  1411} 121 (\textit{Preprint} \eprint{1408.0728})

\bibitem{Buras:2015yca}
Buras A~J, Buttazzo D and Knegjens R 2015 {\em JHEP\/} {\bf 11} 166
  (\textit{Preprint} \eprint{1507.08672})

\bibitem{Buras:2015jaq}
Buras A~J 2016 {\em JHEP\/} {\bf 04} 071 (\textit{Preprint}
  \eprint{1601.00005})

\bibitem{Buras:2012jb}
Buras A~J, De~Fazio F and Girrbach J 2013 {\em JHEP\/} {\bf 1302} 116
  (\textit{Preprint} \eprint{1211.1896})

\bibitem{Grossman:1997sk}
Grossman Y and Nir Y 1997 {\em Phys.~Lett.\/} {\bf B398} 163--168
  (\textit{Preprint} \eprint{hep-ph/9701313})

\bibitem{Blanke:2015wba}
Blanke M, Buras A~J and Recksiegel S 2016 {\em Eur. Phys. J.\/} {\bf C76} 182
  (\textit{Preprint} \eprint{1507.06316})

\bibitem{Buras:2015kwd}
Buras A~J and De~Fazio F 2016 {\em JHEP\/} {\bf 03} 010 (\textit{Preprint}
  \eprint{1512.02869})

\bibitem{Buras:2016dxz}
Buras A~J and De~Fazio F 2016 {\em JHEP\/} {\bf 08} 115 (\textit{Preprint}
  \eprint{1604.02344})

\bibitem{Tanimoto:2016yfy}
Tanimoto M and Yamamoto K 2016  (\textit{Preprint} \eprint{1603.07960})

\bibitem{Kitahara:2016otd}
Kitahara T, Nierste U and Tremper P 2016  (\textit{Preprint}
  \eprint{1604.07400})

\bibitem{Endo:2016aws}
Endo M, Mishima S, Ueda D and Yamamoto K 2016  (\textit{Preprint}
  \eprint{1608.01444})

\bibitem{Singer:1980sw}
Singer M, Valle J~W~F and Schechter J 1980 {\em Phys. Rev.\/} {\bf D22} 738

\bibitem{Pisano:1991ee}
Pisano F and Pleitez V 1992 {\em Phys.~Rev.\/} {\bf D46} 410--417
  (\textit{Preprint} \eprint{hep-ph/9206242})

\bibitem{Frampton:1992wt}
Frampton P~H 1992 {\em Phys.~Rev.~Lett.\/} {\bf 69} 2889--2891

\bibitem{Foot:1992rh}
Foot R, Hernandez O~F, Pisano F and Pleitez V 1993 {\em Phys. Rev.\/} {\bf D47}
  4158--4161 (\textit{Preprint} \eprint{hep-ph/9207264})

\bibitem{Montero:1992jk}
Montero J~C, Pisano F and Pleitez V 1993 {\em Phys. Rev.\/} {\bf D47}
  2918--2929 (\textit{Preprint} \eprint{hep-ph/9212271})

\bibitem{Buras:2012dp}
Buras A~J, De~Fazio F, Girrbach J and Carlucci M~V 2013 {\em JHEP\/} {\bf 1302}
  023 (\textit{Preprint} \eprint{1211.1237})

\bibitem{Buras:2013dea}
Buras A~J, De~Fazio F and Girrbach J 2014 {\em JHEP\/} {\bf 1402} 112
  (\textit{Preprint} \eprint{1311.6729})

\bibitem{Buras:2014yna}
Buras A~J, De~Fazio F and Girrbach-Noe J 2014 {\em JHEP\/} {\bf 1408} 039
  (\textit{Preprint} \eprint{1405.3850})

\bibitem{Diaz:2004fs}
Diaz R~A, Martinez R and Ochoa F 2005 {\em Phys.~Rev.\/} {\bf D72} 035018
  (\textit{Preprint} \eprint{hep-ph/0411263})

\bibitem{CarcamoHernandez:2005ka}
Carcamo~Hernandez A, Martinez R and Ochoa F 2006 {\em Phys.~Rev.\/} {\bf D73}
  035007 (\textit{Preprint} \eprint{hep-ph/0510421})

\bibitem{Aaboud:2016ire}
Aaboud M {\em et~al.\/} (ATLAS) 2016  (\textit{Preprint} \eprint{1604.04263})

\bibitem{Altmannshofer:2014rta}
Altmannshofer W and Straub D~M 2015 {\em Eur. Phys. J.\/} {\bf C75} 382
  (\textit{Preprint} \eprint{1411.3161})

\bibitem{Descotes-Genon:2015uva}
Descotes-Genon S, Hofer L, Matias J and Virto J 2015  (\textit{Preprint}
  \eprint{1510.04239})

\bibitem{Bobeth:2016llm}
Bobeth C, Buras A~J, Celis A and Jung M 2016  (\textit{Preprint}
  \eprint{1609.04783})

\bibitem{Ishiwata:2015cga}
Ishiwata K, Ligeti Z and Wise M~B 2015 {\em JHEP\/} {\bf 10} 027
  (\textit{Preprint} \eprint{1506.03484})

\bibitem{Altmannshofer:2014cfa}
Altmannshofer W, Gori S, Pospelov M and Yavin I 2014 {\em Phys. Rev.\/} {\bf
  D89} 095033 (\textit{Preprint} \eprint{1403.1269})

\end{thebibliography}
\end{document}